\documentclass[aps,prd,twocolumn,twoside,superscriptaddress,floatfix]{revtex4}
\usepackage{amsmath}
\begin{document}
\def\d0{\delta_0}
\def\dL{\delta_L}

\newcommand\beq{\begin{equation}}
\newcommand\eeq{\end{equation}}
\newcommand{\be}{\begin{equation}}
\newcommand{\ee}{\end{equation}}
\newcommand\eea{\end{eqnarray}}
\newcommand\bea{\begin{eqnarray}}
\newcommand{\bk}{{\bf k}}
\newcommand{\bq}{{\bf q}}
\newcommand{\bx}{{\bf x}}
\newcommand{\bd}{{\bf d}}
\newcommand{\br}{{\bf r}}

\newcommand{\aaps}{{Astron.~Astrophys.~Supp.}}

\newcommand{\jcap}{{J.~Cosm.~Astrop.~Phys.}}
\newcommand{\araa}{{Annu.~Rev.~Astron.~Astrophys.}}
\newcommand{\aap}{{Astron.~Astrophys.}}
\newcommand{\apjl}{{Astrophys.~J.~Lett.}}
\newcommand{\apjs}{{Astrophys.~J.~Supp.}}
\newcommand{\aj}{{Astron.~J.}}
\newcommand{\mnras}{{Mon.~Not.~R.~Astron.~Soc.}}

\def\la{\langle}
\def\ra{\rangle}

\newcommand{\lexp}{\mathop{\langle}}
\newcommand{\rexp}{\mathop{\rangle}}
\newcommand{\rexpc}{\mathop{\rangle_c}}
\newcommand{\beqa}
{\begin{eqnarray}}
\newcommand{\eeqa}{\end{eqnarray}}

\def\k{{\hbox{\bf k}}}
\def\q{{\hbox{\bf q}}}
\def\x{{\hbox{\bf x}}}
\def\r{{\hbox{\bf r}}}
\def\v{{\hbox{\bf v}}}
\def\u{{\hbox{\bf u}}}
\def\dD{\delta_{\rm D}} 
\def\tvk{{\hat{\k}}}
\def\tvq{{\hat{\q}}}

\def\Dp{D_+}
\def\de{\delta}
 
\def\Mpc{\, h^{-1} \, {\rm Mpc}}
\def\Gpc{\, h^{-1} \, {\rm Gpc}}
\def\Gpccube{\, h^{-3} \, {\rm Gpc}^3}
\def\kvecMpc{\, h \, {\rm Mpc}^{-1}}
\def\ga{\mathrel{\mathpalette\fun >}}
\def\fun#1#2{\lower3.6pt\vbox{\baselineskip0pt\lineskip.9pt
        \ialign{$\mathsurround=0pt#1\hfill##\hfil$\crcr#2\crcr\sim\crcr}}}

\title{The Shift of the Baryon Acoustic Oscillation Scale: A Simple Physical Picture}
\begin{abstract}
A shift of the baryon acoustic oscillation (BAO) scale to smaller values than predicted by linear theory was observed in simulations. In this paper, we try to provide an intuitive physical understanding of why this shift occurs, explaining in more pedagogical detail earlier perturbation theory calculations. We find that the shift is mainly due to the following physical effect. A measurement of the BAO scale is more sensitive to regions with long wavelength overdensities than underdensities, because (due to non-linear growth and bias) these overdense regions contain larger fluctuations and more tracers and hence contribute more to the total correlation function.  In overdense regions the BAO scale shrinks because such regions locally behave as positively curved closed universes, and hence a smaller scale than predicted by linear theory is measured in the total correlation function. Other effects which also contribute to the shift are briefly discussed. We provide approximate analytic expressions for the non-linear shift including a brief discussion of biased tracers and explain why reconstruction should entirely reverse the shift. Our expressions and findings are in agreement with simulation results, and confirm that non-linear shifts should not be problematic for next-generation BAO measurements.

\end{abstract}
\author{Blake D.~Sherwin}
\affiliation{Department of Physics, Princeton University, Princeton, NJ 08544, USA}
\author{Matias Zaldarriaga}
\affiliation{School of Natural Sciences, Institute for Advanced Study, Princeton, NJ 08540, USA}
\maketitle
\section{Introduction}
The primordial photon-baryon plasma supports the propagation of sound waves, sourced by the initial density fluctuations, until the photons and baryons decouple. At decoupling, a characteristic correlation length is hence imprinted into both the baryons and the photons: the sound horizon at recombination. This manifests itself in the large scale structure of the universe as an excess probability of finding pairs of galaxies (or other tracers) at a separation equal to the sound horizon scale, known in this context as the Baryon Acoustic Oscillation (BAO) scale.  As the value of the sound horizon is well known from microwave background observations, the BAO scale (which is $\sim150$ comoving Mpc) can be used as a ``standard ruler'' to infer the expansion history of the universe. It has been detected (either as a peak in the real space correlation function $\xi(r)$ or as a series of peaks in its Fourier transform, the power spectrum $P(k)$) by a number of large scale structure surveys \citep{Eis05,Col05,Hut06, Teg06, Per07, Bla07,Pad07, Oku08, San09,Kaz10,Per10,Seo12,Ho12,Xu12,Meh12,Pad12}, but its potential as a cosmological probe is nowhere near exhausted: due to its robustness to systematic errors, many future large scale structure surveys propose to measure the BAO scale to $\sim0.1\%$ in order to infer the expansion history of the universe and the properties of dark energy to high precision, e.g.\ \cite{Sch11}. This proposed future research program relies on the BAO scale being a good standard ruler to the sub-percent level.

However, the extent to which this is true needs to be examined carefully, as the BAO scale is somewhat modified by non-linear gravitational effects. Simulations of both dark matter and biased tracers \citep{Seo10, Meh11} show that the BAO standard ruler shrinks by $\sim0.3\% \left[ D(z)/D(0)\right ]^2$ at redshift $z$, where $D(z)$ is the growth factor. Approaches using perturbation theory for both matter and biased tracers \citep{Smi08, Croc08,Pad09} also find a shift of this order, which is explained as being due to the generation of out-of-phase oscillations due to non-linear mode coupling effects. In this paper, we try to develop a more intuitive physical picture for why the BAO scale shrinks due to nonlinearities. We first examine the effect of a long wavelength mode on the small scale correlation function and heuristically derive an approximate expression for the shift in the BAO scale. We then present a less approximate pedagogical calculation of the shift with perturbation theory. In the final sections, we discuss the effect of biased tracers and reconstruction on the shift of the BAO scale.

\section{The physical origin of the BAO scale shift}
In this section, we will describe heuristically how the correlation function on small scales $\xi_S(r)$ (which contains a BAO-like feature) is affected by the presence of a mode  $\delta_L$ with a wavelength larger than the scales of interest which are of order the BAO scale. We will obtain through this simple argument an expression for the shift in the BAO scale. In the following section, we will confirm and refine this expression using perturbation theory.
 
As we are only interested in the correlation function on scales of order the BAO scale $r_B$ (and not much larger), we split the overdensity $\delta$ into long wavelength perturbations $\delta_L$ and short wavelength perturbations $\delta_S$ so that $\delta = \delta_S + \delta_L$. The small scale correlation function can hence be defined as
\beq
\xi_S(r) \equiv \la \delta_S(\mathbf{x}+\mathbf{r}) \delta_S(\mathbf{x})\ra, 
\eeq

We assume that the small scale correlation function is determined with a fixed background overdensity $\delta_L$ in a small volume (smaller than the wavelength of the long mode, but larger than the BAO scale). As shown in \cite{Bal11} and as can be deduced from Birkhoff's theorem, physics in such a small volume is indistinguishable (to first order in $\delta_L$) from physics in a universe with the same curvature as that induced locally by the overdensity $\delta_L$ (and with the same early time limiting behavior as our flat universe). Hence the behavior of small scale density perturbations in the presence of $\delta_L$ is identical to their behavior in an appropriately curved universe.

What are the effects of $\delta_L$, and hence background curvature, on the small scale correlation function $\xi_S$? When $\delta_L$ is positive in a volume, the volume behaves locally as slightly closed universe with positive curvature. The first effect of this is that all the features in the correlation function are contracted to smaller scales than in a flat universe by a factor $\Gamma \equiv a_\mathrm{curved}(\delta_L)/a_0$, where $a$ is the scale factor. As the overall long wavelength density in this region must be simply related to $\Gamma$ by $1+\delta_L=({1\over \Gamma})^3$, the linear theory correlation function $\xi_{S0}(r)$ becomes $\xi_{S0}(r/\Gamma(\delta_L))\approx\xi_{S0}((1 + {\delta_L \over3})r) $ due to this non-linear contraction (to first order in $\delta_L$, neglecting a second order term). 

The second effect of a long wavelength overdensity is on the growth of small scale perturbations. Short modes in regions where $\delta_L$ is positive experience more growth than modes in regions where $\delta_L$ is negative. This can again be explained as overdense regions locally behaving like a closed universe, in which the growth of local perturbations (i.e. the growth of perturbations defined with respect to the background density of the closed universe $\rho_l = \bar{\rho}+\delta \rho_L$) is enhanced by a factor $1+13/21\delta_L$ (see e.g.~\cite{Bal11}) and hence the local correlation function is enhanced by $(1+13/21\delta_L)^2$. To convert a local correlation function defined with respect to $\rho_l$ to one defined globally with respect to $\bar{\rho}$, we must multiply it by $(1+\delta_L)^2$. Hence the growth factor of the short wavelength correlation function due to the presence of a long wavelength overdensity is $(1+{13\over21}\delta_L)^2(1+\delta_L)^2\approx(1+{68\over21}\delta_L)$ (the local curved universe growth factor multiplied by a local-global conversion factor).

Combining these two effects -- non-linear contraction and growth -- we can write down the correlation function on small scales in the presence of a long wavelength mode $\delta_L$

\bea
\label{simpleF}
\xi_S(r) &\approx& (1+{68\over21}\delta_L)\xi_{S0}((1 + {\delta_L \over3})r) \nonumber \\
&\approx& \xi_{S0}(r)+({68 \over 21} \xi_{S0}(r)+{1 \over 3}r\xi'_{S0}(r))\delta_L\\&+&\left[{68 \delta_L\over21} {\delta_L\over3}r\xi'_{S0}(r)\right] + \dots \nonumber
\eea
where we denote linear quantities with subscript zeros and we have neglected other second order terms as well as other effects.

Does the presence of long wavelength modes shift the BAO feature to a different average scale? To find the shift predicted from our expression (\ref{simpleF}), we must compare our expression with a shifted correlation function $\xi(r)=\xi^0(\alpha r)$. Expanding for small shift values $\alpha-1$ gives
\beq
\xi_{S0}(\alpha r) \approx \xi_{S0} + (\alpha-1)r\xi'_{S0}(r)
\eeq
Though in measurements with real and simulated data the shift $\alpha-1$ is typically obtained from a least-squares fit of a shifted template correlation function to the data, we can obtain an approximate estimate of the shift which best fits our expression (\ref{simpleF}) by isolating the coefficient of $r\xi'_{S0}(r)$ after averaging over long wavelength modes. Isolating this coefficient should give approximately similar answers to fitting for the shift using a template which is either linear or has only been changed in amplitude by non-linear effects, as in \cite{Pad09}. The lowest order non-zero contribution to this coefficient is of order $\delta_L^2$:
\beq
\boxed{\alpha-1 \sim \frac{68}{63} \la \delta_L^2\ra}
\eeq
We can see that in this simplified picture (of equation \ref{simpleF}), the BAO scale shift arises from the following effect: \emph{Because perturbations in regions with long-wavelength overdensities grow larger, such regions contribute more to the correlation function, and so we tend to predominantly observe overdense regions when measuring the BAO scale from the correlation function. In overdense regions, the BAO scale shrinks because such regions locally behave as positively curved closed universes, so that we measure a smaller scale than predicted by linear theory.} If we measure the correlation function using biased tracers, the effect should be larger, as we are even more sensitive to overdense regions and hence should see an even larger shift.
We note that our explanation can also be phrased in Newtonian terms instead of a ``local-closed-universe'' picture: Perturbations in overdense regions contribute more to the correlation function due to nonlinear growth, and one hence observes a net compression of the correlation function because such overdense regions experience gravitational infall.

Our derivation assumes that the BAO shift is due to the influence of long wavelength modes. Though this makes sense physically (short wavelength modes fluctuate rapidly over distances smaller than the BAO scale and hence should not lead to coherent gravitational shifts of it), we examine this assumption in the following section. Furthermore, our simple model neglects the influence of anisotropies (it assumes spherical symmetry) as well as other small corrections and neglects other possible second or higher order effects. A perturbation theory calculation including all these effects, leading to a slightly modified result, is shown in the following section. However, most of the shift can be described by the simple physical picture just presented.

The simple arguments of the previous section not only allow a derivation of the shift, but can also provide an equation for the squeezed bispectrum. The squeezed bispectrum is the limit of the three point correlation function in which one of the modes has a much longer wavelength than the other two. Multiplying our heuristic expression (\ref{simpleF}) by $\delta_L$ and averaging over realizations of the long wavelength modes we obtain the following expression for the squeezed bispectrum:

\beq
\la \delta_L \xi_S(r)\ra \sim \left[{68 \over 21} \xi_{S0}(r)+{1\over3} r\xi'_{S0} \right ]\la \delta^2_L \ra
\eeq
where $\delta^2_L$ is the variance of the long wavelength mode.

It should be noted that in simulations, the BAO peak is not only found to be shifted but also to be broadened by non-linear effects. Though one could gain insight into the peak broadening with similar arguments to those used to explain the shift in the previous section, such a discussion is beyond the scope of this paper and we leave it to future work.

\section{Perturbation theory calculation of the BAO scale shift}
In this section we will use second order Eulerian perturbation theory to derive a more rigorous expression for the shift and compare it with our heuristically derived expression. Though other calculations using the machinery of renormalized perturbation theory and Lagrangian perturbation theory have been performed in \cite{Croc08,Pad09}, a simple Eulerian approach is more pedagogical and still captures most of the physics (see also the similar discussion of the largest of the shift terms in \cite{Croc08}).  

In standard perturbation theory we expand the density field in powers of the primordial perturbation $\delta_0$:
\be
\delta=\delta^{(1)} + \delta^{(2)} + \delta^{(3)} + \cdots
\ee
where $\delta^{(n)}$ contains $\d0$ to the $n$th power. 

We will be interested in the power spectrum on the scale of the BAO feature (``small scales'' for our purposes) as well as the squeezed limit of the three point function. The small scale power spectrum has two contributions, $\la \delta^{(2)} \delta^{(2)}\ra $ and  $\la \delta^{(1)} \delta^{(3)}\ra$; however, the 1-3 term can easily be shown to be a simple rescaling of the linear power spectrum and hence does not contribute to the shift. Furthermore, as the three point function involves $\la \delta  \delta^{(1)} \delta^{(1)} \ra$ to lowest order, we only need to know $\delta$ up to second order to evaluate its leading order contribution. Hence to determine both the shift in the small scale power spectrum and the squeezed Bispectrum we only need to study the structure of $\delta^{(2)}$. 

In Fourier space $\delta^{(2)}$ is given by (see e.g.~\cite{Mac09})
\be
\delta^{(2)}(\bk)= \int {d^3 q \over (2\pi)^3} F_2(\bk-\bq,\bq) \d0(\bk-\bq)\d0(\bq) \ee
where
\be
F_2(\bk,\bq)= {17 \over 21} + {1 \over 2} {\bk\cdot\bq \over k q}\  ({k\over q} + {q\over k}) + {2\over 7} [({\bk\cdot\bq \over k q})^2 -{1\over 3}].
\ee

We will initially work in real space for clarity, where the equivalent expression is: 
\be
\delta(\bx)= \d0(\bx)+\underbrace{\mathbf{d}(\bx)\cdot \nabla \d0(\bx)}_\text{shift} + \underbrace{{17\over 21} \d0^2(\bx)}_\text{growth} + \underbrace{{2\over 7} K_{ij}(\bx) K_{ij}(\bx)}_\text{anisotropy} + \cdots, 
\ee
where 
\bea
\mathbf{d}(\bx)& =& - \int {d^3 q \over (2\pi)^3} {i \mathbf{q} \over q^2} \delta_0(\bq) e^{i \bq\cdot\bx} \nonumber \\
K_{ij}(\bx)& =&  \int {d^3 q \over (2\pi)^3} ({ q_i q_j\over q^2}-{1 \over 3}) \delta_0(\bq)  e^{i \bq\cdot\bx}.
\eea
where indices $i,j$ describe projections along three orthonormal coordinate axes and we assume the usual Einstein summation convention.

Here the shift term encodes the motion of the density perturbations due to the gravitational potential sourced by other perturbations, the growth term describes the non-linear growth of perturbations, and the anisotropy term describes the anisotropic distortion of the perturbations.

Hence we obtain to fourth order in $\d0$
\bea
\delta(\mathbf{x}_1)\delta(\mathbf{x}_2) &=&\left[{17\over 21} \nabla \d0(\bx_1)\cdot \bd(\bx_1)\d0^2(\bx_2) +(1 \leftrightarrow 2)\right] \nonumber \\
&+&\left[{2\over 7} \nabla \d0(\bx_1)\cdot \bd(\bx_1)K_{ij}(\bx_2) K_{ij}(\bx_2) +(1 \leftrightarrow 2)\right] \nonumber \\
&+& \nabla \d0(\bx_1)\cdot \bd(\bx_1)\nabla \d0(\bx_2)\cdot \bd(\bx_2) \nonumber \\
&+&\left[{17\over 21}{2\over 7} \d0^2(\bx_1)K_{ij}(\bx_2) K_{ij}(\bx_2) +(1 \leftrightarrow 2)\right]\nonumber \\
&+& \left({2\over 7}\right)^2  K_{ij}(\bx_2) K_{ij}(\bx_2)K_{ij}(\bx_1) K_{ij}(\bx_1)  \nonumber \\
&+&\left({17\over 21}\right)^2\la
\d0^2(\bx_1)\d0^2(\bx_2)\ra \nonumber \\&+&\left[\d0(\bx_1)\delta(\bx_2) +(1 \leftrightarrow 2)\right] +\cdots\nonumber \\ 
\eea
where we have neglected the 1-3 term which does not contribute to the shift.

We now take the average of this quantity so that we can derive the shift by identifying the coefficient of $r \xi'(r)$. To begin, we consider the first term from the equation above (the growth-shift term), at a radius of order the BAO scale $r_B$:
\bea
&&\la{17\over 21} \nabla \d0(\bx_1)\cdot \bd(\bx_1)\d0^2(\bx_2) +(1 \leftrightarrow 2)\ra  \\ &=&-{68\over 21}\nabla_r \xi_0(r) \cdot \nabla_r \int {d^3 q \over (2\pi)^3} {P_0(q) \over q^2} \cos\left[ \bq\cdot (\bx_1-\bx_2) \right] \nonumber 
\eea
We can see that for modes with wavenumber $q>>r\equiv |\bx_1-\bx_2|\sim r_B$, the integrand above is highly oscillatory and averages to zero. This makes physical sense, because short wavelength modes cannot coherently shrink the correlation function on scales $\sim r_B$ much larger than their wavelength, as their gravitational effects average to zero over the BAO scale. Hence one can cut off the integrand for wavelengths above $q>\Lambda$, where $\Lambda \sim 1/r_B$, and expand for $q r <1$, so that the integral simply becomes an average over long wavelength modes (longer than $r_B$):

\bea&&\la{17\over 21} \nabla \d0(\bx_1)\cdot \bd(\bx_1)\d0^2(\bx_2) +(1 \leftrightarrow 2)\ra  \\  &\approx& {68\over 21} \xi_0'(r) {\partial \over \partial r} \int {d^3 q \over (2\pi)^3} {P_0(q) \over q^2} (q^2 r^2 \mu^2/2) \nonumber \\ &\approx&  \left[ {68\over 63}\int_0^\Lambda {dq \over (2\pi)^3} 4 \pi q^2 P_0(q) \right ] r \xi'_{S0}(r) \equiv \left[ {68 \over 63 } \la \delta^2_L \ra \right] r\xi_{S0}'(r) \nonumber
\eea
where $\mu = \hat{\bq} \cdot \hat{\mathbf{r}}$.
Though there is still a contribution to the integral from scales somewhat smaller than the cutoff $\Lambda=1/r_B$, this contribution does not scale as $r$ and hence the term associated with it does not have the form of a shift of the correlation function $\sim r \zeta'(r)$.   Partially due to the approximate nature of the cutoff choice, it should be noted that our expressions for the shift can only be order of magnitude estimates.

All the other terms are similarly oscillatory (see Appendix \ref{oscil}, where they are all written in terms of Bessel functions) so that they can be similarly cut off. We thus obtain for each of the terms (ignoring a uniform shift independent of position as well as the 1-3 term):
\bea
\xi_{S}(r)&=&\la \delta_S(\mathbf{x}_1)\delta_S(\mathbf{x}_2) \ra =
\left[ {68 \over 63 }  r\xi'_{S0}(r) \right]\la \delta^2_L \ra \nonumber \\
&+&\left[ {32 \over 315 }  r\xi'_{S0}(r) \right]\la \delta^2_L \ra \nonumber \\
&+&\left[{r\xi'_{S0}(r)\over15} + {r^2\xi''_{S0}(r)\over10} + {2\xi_{S0}(r)\over3}\right] \la \delta^2_L \ra \nonumber \\
&+&0\nonumber \\
&+& \left({2\over 7}\right)^2\left(16\over 45\right)\xi_{S0}(r) \la \delta^2_L \ra \nonumber \\
&+&\left({34\over 21}\right)^2\xi_{S0}(r) \la \delta^2_L\ra \nonumber \\ 
&+&\xi_{S0}(r)+\cdots \
\eea
or succinctly
\bea
\label{avxi}
 \xi_S(r) &=& \xi_{S0}(r) \nonumber\\&&+ [{2438\over 735} \xi_{S0}(r) + {131\over 105} r \xi^{\prime}_{S0}(r) +{1\over 10} r^2 \xi^{\prime\prime}_{S0}(r)  ] \la\delta_L^2\ra \nonumber\\&&+\cdots
\eea
Note that this expression can be obtained more directly by writing $\delta =\delta^S+\delta^L$ and keeping the long contributions to $d$, $\delta$ and $K$ but not to the derivative of $\delta$ to give the expression
\be
\delta_S(\bx)= \delta_{S0}(\bx+\bd^L)+{34\over 21} \delta_{L0} \delta_{S0}(\bx) + {4\over 7} K^S_{ij}(\bx) K^L_{ij}(\bx) +\cdots, 
\ee
and calculating the short mode correlation function $\xi_S(r)=\la \delta_S \delta_S \ra$.

As before, we identify the coefficient of  $r \xi^{\prime}_{S0}(r) $ in equation (\ref{avxi}) as the shift of the real space correlation function $\alpha-1$:
\beq
\boxed{\alpha-1 = \frac{131}{105} \la \delta_L^2\ra}
\eeq

The $131/105$ coefficient in front of the shift term is $131/105=68/63+32/315+1/15$, where the first term arises from the cross between the shift and growth terms, the second term arises from the cross between the anisotropic and shift terms and the third term comes from the square of the shift term alone. The first term, which is what we obtained from our simple physical argument, is the dominant term (the other two terms are less than 10\% of its magnitude).

In Fourier space our expression leads to: 
\bea
P_S(k)&=&P_{S0}(k)\nonumber \\&+&\left[{569 \over 735} P_{S0}(k) - {47 \over 105} k P_{S0}^\prime(k) +   {1 \over 10} k^2 P_{S0}^{\prime\prime}(k)\right]  \la \delta_L^2\ra \ \nonumber\\\ &+&\cdots
\eea
where we have used
\be
r \xi^{\prime}(r) \rightarrow - {\partial \over \partial k_a} [k_a P(k)] = -(k P^\prime(k) + 3 P(k)).
\ee 
and
\bea
r^2 \xi^{\prime\prime}(r) = r {d \over dr} \left[  r {d \over dr}   \xi(r)\right] - r \xi^{\prime}(r),\nonumber\\
 \rightarrow   {\partial \over \partial k_a} \left[ k_a  {\partial \over \partial k_b}\left[ k_b P(k)\right] \right]  + {\partial \over \partial k_a} \left[ k_a P(k) \right] \nonumber \\ \rightarrow 12P(k) + 8 k P'(k) + k^2 P''(k).
\eea
The direct calculation in Fourier space (expanding such that one mode $q$ has a much longer wavelength than the other, $k>>q$) gives:
\bea
&&P_S(k) \approx P_{S0}(k)\nonumber \\&&+\int  {d^3 q \over (2\pi)^3} 4 [F_2(\bk-\bq,\bq)]^2 P_0(|\bk-\bq|)P_0(q)+\cdots\nonumber\\&&  \approx P_{S0}(k)+ \left[{569 \over 735} P_{S0}(k) - {47 \over 105} k P_{S0}^\prime(k) +   {1 \over 10} k^2 P_{S0}^{\prime\prime}(k)\right] \la \delta_L^2\ra \nonumber \\&&+\int  {d^3 q \over (2\pi)^3} P_0(q){1\over 3} {k^2 \over q^2} P_0(k)+\cdots
\eea
which is the same as our previous expression except for the ${k^2/3 q^2} P(k)$ term, which comes from the uniform shift we ignored in the real space calculation. This shift moves both points in the correlation function by the same amount so it leads to no effect. In the standard perturbation theory calculation this term cancels when $P_{22}$ is combined with $P_{13}$ (see Appendix \ref{third}).

The Fourier space shift $\alpha_k$ describes how much the linear spectrum is rescaled as $P_{S0}(k/\alpha_k)$. Expanding in $\alpha_k-1$ as before, we find we can determine it by identifying the coefficient of the shift term $-kP_{S0}'(k)$
\beq
\label{kspaceform}
\boxed{\alpha_k-1 = \frac{47}{105} \la \delta_L^2\ra}
\eeq
Note that the shift as calculated in Fourier space is different from the shift in the real space correlation function. This is due to the the second derivative terms mixing ``shift'' and ``non-shift'' terms upon Fourier transforming; physically, there is a change of BAO peak shape which looks like a shift in Fourier space for our simple estimation procedure. In general, as the shift depends somewhat on the estimator and is not an exact physical quantity that is conserved under Fourier transformation, one should not expect to get exactly the same result in real and Fourier space if the shape of the correlation function is modified beyond a pure rescaling by non-linear effects. Our estimation of the shifts is only approximate as mentioned previously; the use of more sophisticated shift fitting procedures (especially those which already include a modified peak shape in the correlation function template, as is done in practice) should substantially reduce any difference between shifts estimated in real and Fourier space.

We can also again calculate the squeezed limit of the three point function. We obtain
\be
\boxed{\la \delta_L\xi_S(r)\ra = [{68\over 21} \xi_{S0}(r) + {1\over 3} r \xi^{\prime}_{S0}(r) ] \la \delta_L^2 \ra.} 
\ee
in exact agreement with our intuitive arguments.
To go to Fourier space we transform $r \xi^{\prime}_{S0}(r)$ to give: 
\be
 [{47\over 21} P_{S0}(k)  - {1\over 3} k P_{S0}^\prime(k) ] \la \delta^2_L \ra. 
\ee
which agrees with a direct Fourier space calculation of the squeezed Bispectrum.

\section{The effects of tracer bias on the measured BAO scale}

In order to compare our calculation with the semianalytical results of \cite{Pad09}, we now include bias in the same way using the simplified expression
\be
\delta_h = b_1 \delta + {1 \over 2}b_2 \delta^2 + \cdots
\ee

The averaged two point function now gives
\bea
\la \xi_{Sh}(r)\ra&=& b_1^2 \xi_{S0}(r) \nonumber \\ &+&\la \delta^2_L\ra  \{ b_1^2 ({2438\over 735} \xi_{S0}(r) + {131\over 105} r \xi^{\prime}_{S0}(r) +{1\over 10} r^2 \xi^{\prime\prime}_{S0}(r)  ) 
\nonumber \\
&&+  b_1 b_2 ({68 \over 21}  \xi_{S0}(r) + {2\over 3}  r \xi^{\prime}_{S0}(r)) + b_2^2 \xi_{S0}(r) \}.
\eea

In Fourier space this gives:
\bea
P_{Sh}(k)&=& b_1^2 P_{S0}(k) \nonumber \\&+&\la\delta^2_L\ra  \{ b_1^2 ({569 \over 735} P_{S0}(k) -{47 \over 105} k P_{S0}^\prime(k) +   {1 \over 10} k^2 P_{S0}^{\prime\prime}(k)) 
\nonumber \\
&+&  b_1 b_2 ({26 \over 21}  P_{S0}(k) - {2\over 3}  k P_{S0}'(k)) + b_2^2 P_{S0}(k) \}.
\eea
We can hence identify the BAO scale shift in Fourier space, including the effects of bias, by finding the coefficient of $-b_1^2 k P'(k)$:
\be
\boxed{\alpha_k-1 = {47 \over 105} ( 1 + {70 \over 47} {b_2 \over b_1})\la \delta^2_L\ra ,}
\ee
As $70/47 =  1.489$, this expression agrees with the numerical calculation in \cite{Pad09} which gave $\alpha_k-1\propto  ( 1 + {1.5} {b_2 \over b_1})$.

Aside from the shift, the squeezed three point function can also be calculated as before
\be
\la \delta^L\xi_{Sh}(r)\ra = (b_1^2 [{68\over 21} \xi_{S0}(r) + {1\over 3} r \xi^{\prime}_{S0}(r) ] + 2 b_1 b_2 \xi_{S0}(r)) \la \delta^2_L \ra. 
\ee
\section{BAO scale shift: comparison with simulations}

Evaluating $ \left< \delta_L^2 \right>$ as the variance at the BAO scale $\sim \sigma^2(140 \mathrm{Mpc})\left[D(z)/D(0)\right]^2$ gives $ \left< \delta_L^2 \right>\approx0.0051\left[D(z)/D(0)\right]^2$ using WMAP cosmological parameters \cite{Dun09}, so that from equation (\ref{kspaceform})

\beq
\boxed{\alpha_k -1 \approx 0.23\% \left[D(z)/D(0)\right]^2 }
\eeq

Considering that the numerical value depends on the choice of cutoff scale and window function, and that we have neglected $\Omega_\Lambda$, this agrees well with the value obtained from simulations in \cite{Seo10} of
\beq
\alpha_k(z)-1 = (0.30\pm0.02)\%\left[D(z)/D(0)\right]^2
\eeq

\section{The effect of reconstruction using the Zel'dovich approximation}
Previous measurements of the BAO scale have employed reconstruction techniques to remove large scale velocity fields which broaden the BAO peak. These methods \cite{Eis07} use the Zel'dovich approximation to attempt to restore the linear density field and hence undo the broadening and degrading of the BAO feature. In this section, we will briefly review the reconstruction procedure and examine to what extent reconstruction also reverses the non-linear shift of the peak.

The Eulerian particle position $\mathbf{x}$ is related to the Lagrangian position $\mathbf{r}$ through the displacement vector $\bq$
\beq
\mathbf{x} = \bq + \mathbf{r}
\eeq
As the small mass elements are related by $\rho_0 d^3r =\rho d^3x$, we can relate the overdensity to the Jacobian and expand to first order:
\beq
{\rho \over \rho_0 }= 1 + \delta = {d^3r \over d^3x} = J^{-1} \approx 1 - \nabla \cdot q
\eeq
So that 
\beq
\delta = -\nabla \cdot q
\eeq
For the linear growing mode, $\bq$ is curl-free so that one can define a potential $\bq = -\nabla \phi_Z$.  In order to only undo large scale velocity flows, one smooths the density field (typically with a Gaussian of width $\sim20$ Mpc) to obtain a new field $\delta^G$ and then solves Poisson's equation for $\phi_Z$ to obtain:
\beq
\phi_Z =  -\int {d^3 q \over (2\pi)^3} {1 \over q^2} \delta^{G}(\bq) e^{i \bq\cdot\bx}\eeq
Hence to undo the large scale velocity flows, one simply displaces the measured density field $\delta$ by $\bq = - \nabla \phi_Z$ where
\beq
\bq = \int {d^3 q \over (2\pi)^3} {i \bq \over q^2} \delta^{G}(\bq) e^{i \bq\cdot\bx}
\eeq
Comparing this with the shift
\beq
\mathbf{d}^L=- \int {d^3 q \over (2\pi)^3} {i\bq \over q^2} \delta_{L0}(\bq) e^{i \bq\cdot\bx}
\eeq
in equation (17), which relates the linear and non-linear fields as
\beq
\delta_S(\bx)= \delta_{S0}(\bx+\bd^L)+{34\over 21} \delta_{L0} \delta_{S0}(\bx) + {4\over 7} K^S_{ij}(\bx) K^L_{ij}(\bx), 
\eeq
we can see that $\bq$ is (very nearly) identical to $-\mathbf{d}_L$ and hence reconstruction (i.e.\ displacing the field by $\bq$) will undo the non-linear shift we found previously, as long as it includes in $\delta^G$ all the long wavelength modes responsible for the shift. As the broadening is due to a wider range of modes than the shift (including shorter wavelengths), if a reconstruction procedure reverses the peak broadening, it automatically also reverses the lowest order shift. Our conclusion that the BAO shift should be greatly reduced after reconstruction agrees with simulation results from \cite{Meh11}, who found that the shifts were consistent with zero after reconstruction was applied. It also agrees with perturbation theory calculations in \cite{Noh09}.
\section{Conclusions}
We have investigated the non-linear shift in the BAO scale found in simulations and have tried to develop an understanding of why the scale becomes smaller due to non-linearities. Using both intuitive arguments and perturbation theory, we found that the main reason for the shift in the BAO scale is as follows: As perturbations in regions with long-wavelength overdensities are larger due to non-linear growth and bias, these regions contribute more to the correlation function so that a measurement of the BAO scale is more sensitive to overdense regions; in overdense regions, the BAO scale shrinks because such regions locally behave as closed universes, so that we measure a smaller scale than predicted by linear theory.

We have expanded and verified these heuristic arguments using perturbation theory, and derived order of magnitude expressions for the real and Fourier space shifts:
\be
\alpha - 1 = {131 \over 105}\la \delta_L^2\ra; \ \ \ \alpha_k - 1 = {47 \over 105}\la \delta_L^2\ra
\ee
where the average is taken over modes with wavelengths longer than the BAO scale (and can be approximated as $\la \delta_L^2\ra\sim  \sigma^2(140 \mathrm{Mpc})\left[D(z)/D(0)\right]^2$). The difference between the shifts in real and Fourier space we estimated approximately should be much smaller for the more sophisticated estimators used in practice. Using a simple model for the shift of biased tracers, we obtained an analytic expression in agreement with \cite{Pad09}. For dark matter our expression for the shift evaluates to $ \alpha_k-1\sim0.23\% \left[D(z)/D(0)\right]^2$, which is in agreement with simulations. We explained in our physical picture why reconstruction of the BAO peak should undo the shifts, as also found in simulations. The sub-percent level of the BAO scale shift and the ease with which it can be undone using reconstruction confirm that it will not be problematic for surveys in the foreseeable future.

\begin{acknowledgements}
We thank Uros Seljak, Marilena LoVerde, Cora Dvorkin and David Spergel for discussions and comments on the draft and thank Nikhil Padmanabhan for discussions. We acknowledge support from a National Science Foundation Graduate Research Fellowship (BDS), NSF grants PHY-0855425,
AST-0506556 \& AST-0907969, and by the David \& Lucile Packard and the John D. \& Catherine T. MacArthur Foundations (MZ).\\

\end{acknowledgements}

\section{Appendix: calculation of the correlation function to third order}
\label{third}
\begin{widetext}

In Fourier space we will write $\delta^{(3)}$ as:
\bea
\delta^{(3)}(\bk)&=& \int {d^3 q_1 \over (2\pi)^3} {d^3 q_2 \over (2\pi)^3}  {d^3 q_3 \over (2\pi)^3} F_3(\bq_1,\bq_2,\bq_3) \times \nonumber \\
&& \d0(\bq_1)\d0(\bq_2) \d0(\bq_2) (2\pi)^3 \delta^D(\bk-(\bq_1+\bq_2+\bq_3)),
\eea
where we have defined
\bea
F_3(\bq_1,\bq_2,\bq_3) &=&  \bar{F}_3(\bq_1,\bq_2,\bq_3)+\bar{F}_3(\bq_2,\bq_1,\bq_3)+\bar{F}_3(\bq_3,\bq_2,\bq_1) \nonumber \\
 \bar{F}_3(\bq_1,\bq_2,\bq_3)&=&{1\over 54} \{ 2 \beta(\bq_1,\bq_2+\bq_3) G_2(\bq_2+\bq_3)+ 7 \alpha(\bq_1,\bq_2+\bq_3) F_2(\bq_2+\bq_3)\nonumber \\
 && + (2 \beta(\bq_2+\bq_3,\bq_1) + 7 \alpha(\bq_2+\bq_3,\bq_1))G_2(\bq_2+\bq_3)\} \nonumber \\
  \alpha(\bq,\bk)&=& {(\bq+\bk)\cdot \bq \over q^2} \nonumber \\
  \beta(\bq,\bk)&=& {|\bq+\bk|^2 \bk\cdot \bq  \over 2 k^2 q^2},
\eea
with
\bea
F_2(\bk,\bq)&=& {17 \over 21} + {1 \over 2} {\bk\cdot\bq \over k q}\  ({k\over q} + {q\over k}) + {2\over 7} [({\bk\cdot\bq \over k q})^2 -{1\over 3}] \nonumber \\
G_2(\bk,\bq)&=& {13 \over 21} + {1 \over 2} {\bk\cdot\bq \over k q}\  ({k\over q} + {q\over k}) + {4\over 7} [({\bk\cdot\bq \over k q})^2 -{1\over 3}].
\eea
These are the standard perturbation theory kernels (see for example \cite{Mac09}, although we have rearranged some of the terms and thus we have introduced $\bar F_3$. In our notation $F_2$, $G_2$ and $F_3$ are the standard symmetrized kernels. 

One can use these kernels to compute the $1-3$ contribution to the power spectrum in the limit $q << k$, 
\bea
\int {d^3 q \over (2\pi)^3} 6\  F_3(\bk,\bq,\bq) P_0(k) P_0(q) &=& \int {d^3 q \over (2\pi)^3} {1\over 21} [ -{21 k^2 \mu^2 \over q^2} +10 - 2 \mu^2 -8 \mu^4]P_0(k) P_0(q) \nonumber \\
&=&    {116 \over 315} P_{S0}(k)\la \delta_L^2\ra-\int  {d^3 q \over (2\pi)^3} P_0(q){1\over 3} {k^2 \over q^2} P_0(k)
\eea

Note that as discussed before the piece proportional to $1/q^2$ cancels in the sum $P_{22}+P_{13}$ to give:
\be
P=P_{22}+P_{13}
=   \{{2519 \over 2205} P_{S0}(k) - {47 \over 105} k P_{S0}^\prime(k) +   {1 \over 10} k^2 P_{S0}^{\prime\prime}(k) \} \la \delta_L^2\ra
\ee

We can also use the above expression for $\delta^{(3)}$ to go directly to real space. We are interested in the effect of a long mode on the short fluctuations and want to use $\delta^{(3)}$ to obtain the piece that is linear in the short modes and quadratic in the long. We obtain:
\bea
\delta^{(3)}(\bx) &= &{1\over 2} d^L_{i}  d^L_{j} \partial_i\partial_j \delta^S(\bx) + A_i  \partial_i \delta^S(\bx) + B  \delta^S(\bx) \nonumber \\
& +&  C_{ij} K_{ij}^S(\bx) + D_{ijk} \partial_i K_{jk}^S(\bx) + E_{ij,kl} K_{ij,kl}^S(\bx) .
\eea
We have introduced $A_i$, $B$, $C_{ij}$, $D_{ijk}$ and $E_{ij,kl}$  which are all quadratic in the long mode and are given by:
\bea
A_i &=& {25 \over 14} \delta^L d^L_i + {1\over 2} d^L_j K^L_{ji} +{1\over 2}  d^{(2)L}_i \nonumber \\
B&=& {682 \over 657} (\delta^L)^2 + {208 \over 189} d^L_i \partial_i \delta^L +  {4\over 27} K_{ij}^L K_{ij}^L  + {7 \over 18} \delta^{(2)L} + {25 \over 54}  \theta^{(2)L}  
\nonumber \\
C_{ij} &=& d^L_i \partial_j \delta^L + {194 \over 189} \delta^L K_{ij}^L + {4 \over 9} K_{ik}^L K_{kj}^L + {2\over 9}  K_{ij}^{(2)L} \nonumber \\
D_{ijk}&=& {4\over 7} d^L_i K_{jk}^L \nonumber \\
E_{ij,kl}&=& - {4 \over 21} K_{ij}^L K_{kl}^L.
\eea
For convenience we have introduced:
\bea
K_{ab,cd}(\bx)& =&  \int {d^3 q \over (2\pi)^3} ({ q_a q_b\over q^2}-{1 \over 3} \delta_{ab})  ({ q_c q_d\over q^2}-{1 \over 3}\delta_{cd}) \delta_0(\bq) e^{i \bq\cdot\bx}. 
\eea
The second order fluctuations are given by:
\bea
\delta^{(2)L}(\bk)&=& \int {d^3 q \over (2\pi)^3} F_2(\bk-\bq,\bq) \d0^L(\bk-\bq)\d0^L(\bq) \nonumber \\ 
\theta^{(2)L}(\bk)&=& \int {d^3 q \over (2\pi)^3} G_2(\bk-\bq,\bq) \d0^L(\bk-\bq)\d0^L(\bq) \nonumber \\ 
d^{(2)L}_a(\bq)&=& -{i q_a \over q^2}  \theta^{(2)L}(\bq)\nonumber \\ 
K_{ab}^{(2)L}(\bq)&=& ({ q_a q_b\over q^2}-{1 \over 3}\delta_{ab}) \d0^{(2)L}(\bq).
\eea
In real space:
\bea
\label{pert2}
\delta^{(2)L}(\bx)&=&d^L_k \partial_k \d0^L+ {17\over 21} (\d0^L)^2 + {2\over 7} K_{ij}^L K_{ij}^L \nonumber \\
\theta^{(2)L}(\bx)&=&d^L_k \partial_k \d0^L+ {13\over 21} (\d0^L)^2 + {4\over 7} K_{ij}^L K_{ij}^L.
\eea

If we are interested in using these expression to compute the monopole of the short scale correlation function, then the terms proportional to $K_{jk}^S$ will not contribute so we can ignore the $C_{ij}$ and $D_{ijk}$ terms. The terms  ${1/2} d^L_{i}  d^L_{j} \partial_i\partial_j \delta^S(\bx) + A_i  \partial_i \delta^S(\bx)$ are responsible for a shift of the small scales that cancels in the correlation function. Thus only the the terms proportional to $B$ and $E_{ij,kl}$ contribute. Note that 
$K_{ij,kl}^S$ cannot be neglected, and when computing averages over angles should be replaced by: 
\be
K_{ij,kl}^S \rightarrow [-{2 \over 45} \delta_{ij} \delta_{kl} + {1\over 15} ( \delta_{ik} \delta_{jl}+ \delta_{il} \delta_{kj})] \delta^S.
\ee

Thus for the purpose of computing the angled averaged correlation function we can use
\bea
\label{d3}
\delta^{(3)}(\bx) \rightarrow [ {682 \over 567} (\delta^L)^2 + {208 \over 189} d^L_i \partial_i \delta^L +  {116\over 945} K_{ij}^L K_{ij}^L  + {7 \over 18} \delta^{(2)L} + {25 \over 27}  \theta^{(2)L} ] \delta^S(\bx)
\eea

When calculating the correlation function, the contribution $\langle \delta^{(3)} \delta + \delta \delta^{(3)} \rangle$ leeds to a term we can call $\xi^S_{13}$ given by:
\bea
\xi^S_{13}(r) &=& 2 \times \langle [ {682 \over 567} (\delta^L)^2  + {208 \over 189} d^L_i \partial_i \delta^L +  {116\over 945} K_{ij}^L K_{ij}^L  + {7 \over 18} \delta^{(2)L} + {25 \over 54}  \theta^{(2)L} ]  \rangle  \xi_0(r) \nonumber \\
&=&2 \times  [  {682 \over 567}  - {208 \over 189} +  {116\over 945} {2 \over 3}   ]  \sigma_L^2  \xi_0(r) =  {116 \over 315}  \sigma_L^2  \xi_0(r) 
\eea 

We can use equation (\ref{pert2}) and (\ref{d3}) to obtain:
\bea
\delta^{(3)}(\bx) \rightarrow [ {58 \over 315} (\delta^L)^2 + {1361 \over 630} \delta^{(2)L} -{131 \over 630}  \theta^{(2)L} ] \delta^S(\bx)
\eea
which when calculating the correlation function gives:
\bea
\xi^S_{13}(r) &=& 2 \times \langle [ {58 \over 315} (\delta^L)^2 + {1361 \over 630} \delta^{(2)L} - {131 \over 630}  \theta^{(2)L} ] \rangle  \xi_{S0}(r) \nonumber \\
&=&2 \times {58 \over 315}   \sigma_L^2  \xi_{S0}(r) =  {116 \over 315}  \sigma_L^2  \xi_{S0}(r) 
\eea 

\section{Appendix: 2-2 Correlation function in real space}
\label{oscil}
The second order expression for the density field can be used to directly obtain the $2-2$ piece of the correlation.
We obtain:
\bea
&&\zeta(r)|_{2-2} = {578 \over 441} \zeta_{0,0}^2 \nonumber \\
&-& {68 \over 21} \zeta_{1,1}   \zeta_{-1,1} \nonumber \\
&+&   {68 \over 49} \zeta_{0,2}^2   \zeta_{-1,1}+ {136 \over 147} \zeta_{0,2}   \zeta_{0,0}+ {68 \over 441} \zeta_{0,0}^2   \nonumber \\
&+&{3 \over 2} \zeta_{2,2}   \zeta_{-2,2}+{1 \over 2} \zeta_{2,2}   \zeta_{-2,0}+{1 \over 2} \zeta_{2,0}   \zeta_{-2,2}+{1 \over 2} \zeta_{2,0}   \zeta_{-2,0}
+{3 \over 2} \zeta_{0,2}^2  + \zeta_{2,0}   \zeta_{0,0}+{1 \over 2} \zeta_{0,0}^2   \nonumber \\
&-& {4 \over 3} \zeta_{1,1}   \zeta_{-1,1}-{12 \over 7} \zeta_{1,3}   \zeta_{-1,1}-{12 \over 7} \zeta_{-1,3}   \zeta_{1,1}-{20 \over 7} \zeta_{1,3}   \zeta_{-1,3} \nonumber \\
&+& {4 \over 7} \zeta_{0,2}^2 +{60 \over 49} \zeta_{0,4}   \zeta_{0,03}+{5 \over 7} \zeta_{0,4}^2 +{20 \over 147} \zeta_{0,2}   \zeta_{0,0}
+{6 \over 49} \zeta_{0,4}  \zeta_{0,0}  +{11 \over 441} \zeta_{0,0}^2.
\eea
The different lines correspond to the different terms in the calculation, growth-growth, growth-shift, growth-anisotropy, shift-shift, shift-anisotropy and anisotropy-anisotropy. 

 We have defined
 \be
  \zeta_{m,n}=\int d\ln k \Delta^2(k) k^m [{\partial \over \partial kr}]^n j_0(kr).
 \ee
Using this notation
\bea
\zeta_{0,0}(r) &=& \zeta(r) \nonumber \\
\zeta_{1,1}(r) &=& \zeta^\prime(r) \nonumber \\
\zeta_{2,2}(r) &=& \zeta^{\prime\prime}(r) \nonumber \\
\zeta_{2,0}(r)&=& - \nabla^2 \zeta(r)=  - \zeta^{\prime\prime}(r) - {2\over r}  \zeta^\prime(r).
\eea

We can now take the limit when one of the momenta is much smaller than $1/r$ which gives:
\bea
{d \zeta(r)|_{2-2} \over d \sigma_L^2}& =& {1156 \over 441} \zeta_{S0}   \nonumber \\
&+& {68 \over 63} r \zeta_{S0}^\prime    \nonumber \\
&+&   0  \nonumber \\
&-&{1 \over 3}  {1 \over q^2} \nabla^2 \zeta_{S0}+{1\over 10} r^2 \zeta_{S0}^{\prime\prime}+{1 \over 15} r \zeta_{S0}^\prime +{2 \over 3} \zeta_{S0}  \nonumber \\
&+& {32 \over 315}r \zeta_{S0}^\prime \nonumber \\
&+& {64 \over 2205} \zeta_{S0},
\eea
which adds to
\be
\la \delta_L^2\ra [{2438 \over 735} \zeta_{S0} +{131 \over 105} r \zeta_{S0}^\prime +{1 \over 10} r^2 \zeta_{S0}^{\prime\prime}-{1 \over 3}  {\sigma_L^2 \over q^2} \nabla^2 \zeta_{S0}].
\ee

\end{widetext}

\end{document}